\documentclass[]{spie}  

\usepackage{float}
\usepackage{xcolor}

\usepackage{amsmath,amsfonts,amssymb}
\usepackage{graphicx}
\usepackage[colorlinks=true, allcolors=blue]{hyperref}
\usepackage{multirow}

\title{Coherence differential imaging using gradient-boosted decision trees for the direct detection of exoplanets}

\author[a]{Johan Mazoyer}
\author[a,b]{María J. Mellado-Tenorio}
\author[a]{Axel Potier}
\author[a]{Christian Wilkinson}
\author[a]{Lukas Delaye}
\author[a]{Raphaël Galicher}
\author[a,c]{Iva Laginja}

\affil[a]{LIRA, Observatoire de Paris, Université PSL, Sorbonne Université, Université Paris Cité, CY Cergy Paris Université, CNRS, 92195 Meudon, France}
\affil[b]{Department of Computer Science (DCC), Universidad de Chile, Santiago, Chile}
\affil[c]{Université Côte d’Azur, Observatoire de la Côte d’Azur, CNRS, Laboratoire Lagrange, France} 

\authorinfo{Send correspondence to johan.mazoyer@obspm.fr}

\begin{document} 
\maketitle

\begin{abstract}
Coronagraphic imaging of exoplanets is limited by residual speckles that mimic planets. Advanced post-processing is essential for current and future instruments on the ground or in space. Current techniques are time-intensive and limited. ADI requires long sequences and is limited at small separations. RDI is also time-consuming and sensitive to speckle evolution, leading to imperfect subtraction. Coherence Differential Imaging (CDI), which we successfully demonstrated on SPHERE, offers a faster alternative by using the light incoherence between speckles and planets. However, its reliance on accurate instrumental models limits its performance. In this work, we introduce EPICX, an enhanced CDI method using gradient-boosted decision trees. EPICX models the differential signal as a high-dimensional regression problem, optimizing the discrimination between coherent speckle noise and incoherent planet signal. We validate this enhanced CDI method using simulated data for different coronagraphs, including those aboard JWST and Roman. 
\end{abstract}


\section{INTRODUCTION}
\label{sec:intro}  

High-contrast imaging detects light from objects like exoplanets or disks in circumstellar environments, enabling spectral, astrometric, and polarimetric characterization. This requires advanced coronagraphs and wavefront sensing and control \cite{galicher_imaging_2024}. Most large ground-based optical telescopes are now equipped with coronographic instruments like VLT/SPHERE, Gemini/GPI, Subaru/SCExAO, and Magellan/MagAO-X. This is also the case for Large Space-based observatory with HST / STIS, JWST / MIRI and Roman / CGI \cite{poberezhskiy_romancgi_2025}. 

However, coronagraphs are severely limited by aberration in the upstream wavefront. For ground-based coronagraph, these aberrations are mostly created by the atmosphere, and mostly corrected by extreme adaptive optics systems. Unfortunately, even the most powerful AO systems do not compensate non-common-path aberrations (NCPA), which result from the difference of the optical path after the beam splitter between the science path and the wavefront sensor path. Similarly, space-based coronagraphic instruments are also sensitive to speckles limitations as their optics also introduce limitations.

Active correction of these speckles is an active research area, with focal plane wavefront sensing techniques being tested and implemented recently on VLT/SPHERE \cite{potier_increasing_2022} and Roman \cite{cady_romanhowfsc_2025}. However, current coronagraphic instruments rely on calibration algorithms in post-processing. Speckles and exoplanet images exhibit distinct behaviors, allowing differential imaging techniques to calibrate speckles and extract astrophysical signals. Angular Differential Imaging (ADI) \cite{marois_angular_2006}, which uses rotation to subtract the stellar speckle pattern, has been widely adopted on the ground. Spectral Differential Imaging (SDI) \cite{racine1999_SDI} and dual-band imaging provide additional calibration but often require combination with ADI for imagers or low-resolution Integral Field Spectrographs (IFS) \cite{flasseur_paco_2020}. However, both methods face critical limitations close to the star ($< 5 \lambda/D$), as even advanced algorithms struggle to detect planets without sufficient planet/speckle diversity \cite{flasseur_paco_2020}.
Reference Differential Imaging (RDI \cite{beuzit1997_RDI}) excels in stable environments, particularly in space \cite{choquet2016_RDI} or through rapid ground-based observations \cite{wahhaj2021_RDI}, but is constrained by speckle evolution during target switching and associated optics movements. RDI’s sensitivity to stellar size and type further complicates its implementation for missions like Roman without impractical slews \cite{hom_romanRDI_2026} and will likely be even more difficult for HWO. Polarization Differential Imaging (PDI) \cite{baba2005_PDI} effectively distinguishes polarized scattered dust light from unpolarized stellar light, making it ideal for disk detections. However, exoplanet polarization remains too faint for detection. Medium- and high-resolution IFS are driving the development of techniques like Molecular Mapping (MM) \cite{Snellen2015_molecularmapping}, which leverages cross-correlation with known planetary spectra to differentiate planetary and stellar light efficiently. However, MM is not applicable to imagers or low-resolution IFS, which are still prevalent in coronagraphic instruments and which still rely mainly on ADI (ground) or RDI (space). 

Building on these observation methods, many algorithms have been developed to leverage the diversity and extract the planetary signal among the speckle, from the most general image combination LOCI \cite{Lafreniere2007_LOCI} and PCA / KLIP \cite{soummer_detection_2012,amara_pynpoint_2012}) to machine-learning-based optimized methods \cite{flasseur_deep_2024, Cantero_NA_SODINN_2023}. For coronagraphic observations, the total exposure time is not primarily determined by the expected flux of the object, as in typical astronomical observations. Instead, it is driven by the need for speckle calibration, such as achieving sufficient parallactic angle rotation for ground-based observations or acquiring a reference image for space-based observations. This also drastically limits their efficiency, as post-processing algorithms are inherently limited by speckle variation during observations.

\begin{figure}[!ht]
    \centering
    \includegraphics[width=0.5\textwidth]{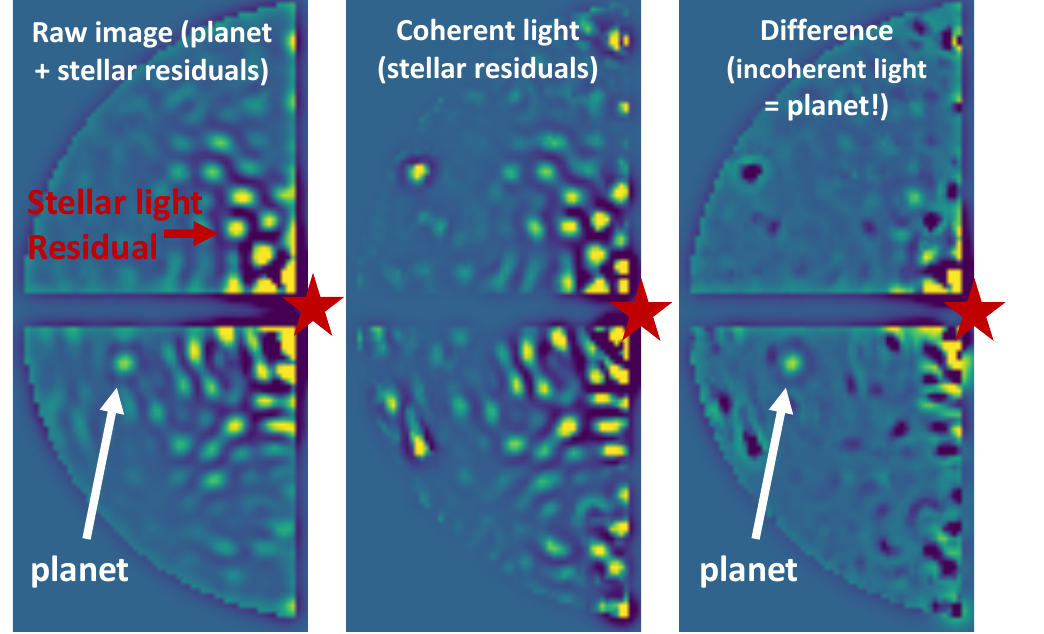}
    \caption{\it CDI based on Pair Wise Probing. \textbf{Left}: Raw SPHERE image. \textbf{Center}: Using a Pair Wise Probing, a diversity-based focal plane wavefront sensing method, we reconstruct a ‘reference’ of the coherent light. \textbf{Right}: Difference between the image and reference: speckles are removed which greatly increase $\beta$-Pictoris b SNR.}
    \label{fig:cdi_betapic} 

\end{figure}

More recently, coherent differential imaging (CDI) techniques have been developed, leveraging focal plane wavefront sensing and control algorithms. These methods rely on the coherent modulation of the speckle field, allowing the coherent starlight signal to be isolated from the incoherent planetary astrophysical signal in post-processing. Mostly based on focal plane wavefront sensing method, these techniques modulate the speckles in the focal plane, either by modifying the coronagraph design to fringe the coherent light (i.e the speckles) \cite{baudoz_laboratory_2013, bottom2017_CDI, gerard_cdiscc_2018}, or by introducing a known diversity using the deformable mirror and recovering the aberrations using a model of the instrument \cite{potier_increasing_2020,nishikawa_coherent_2022}. Some of these techniques have already been demonstrated on-sky to detect substellar companions or circumstellar disks, ar the Palomar Observatory \cite{bottom2017_CDI} or using VLT/SPHERE \cite{potier_coherent_2025}. Fig.~\ref{fig:cdi_betapic} shows previously unpublished on-sky detection of $\beta$-Pictoris b demonstrating the use of CDI on VLT/SPHERE. However, we can see that some speckles remain, in particular close to the axis, limiting the detection capabilities. 

The next generation of coronagraphic instrument, VLT/SPHERE+ \cite{Mazoyer2026SAXOplusNCPA} and Roman CGI \cite{zhou_roman_2020}, will include active speckles correction, relying on the Pair Wise Probing (PWP) wavefront sensing method (see next section). We can therefore use the same estimation method as a CDI post-processing method, relaxing the constraint on the slow active methods and / or limiting the use of long ADI or RDI sequences. 

\section{Methods}

In this study, we base the observation strategy on classical diversity-based focal-plane wavefront sensing, PWP, but use a machine learning decision trees algorithm to separate planet and speckle signals. Unlike the CDI-PWP method shown in Fig~\ref{fig:cdi_betapic}, we do not attempt to measure a reference image for subtraction. We assume that a sequence of probed images contains enough information to classify all PSF-shaped structures in the image as speckles or planets. Thus, it is not currently optimized to detect extended features (e.g., protoplanetary disks) in images.

In this section, we briefly review the principles of PWP, then describe the simulation parameters used to train our dataset and finally the proposed algorithm.

\subsection{Pair Wise Probing}

\begin{figure}[!ht]
    \centering
    \includegraphics[width=0.5\textwidth]{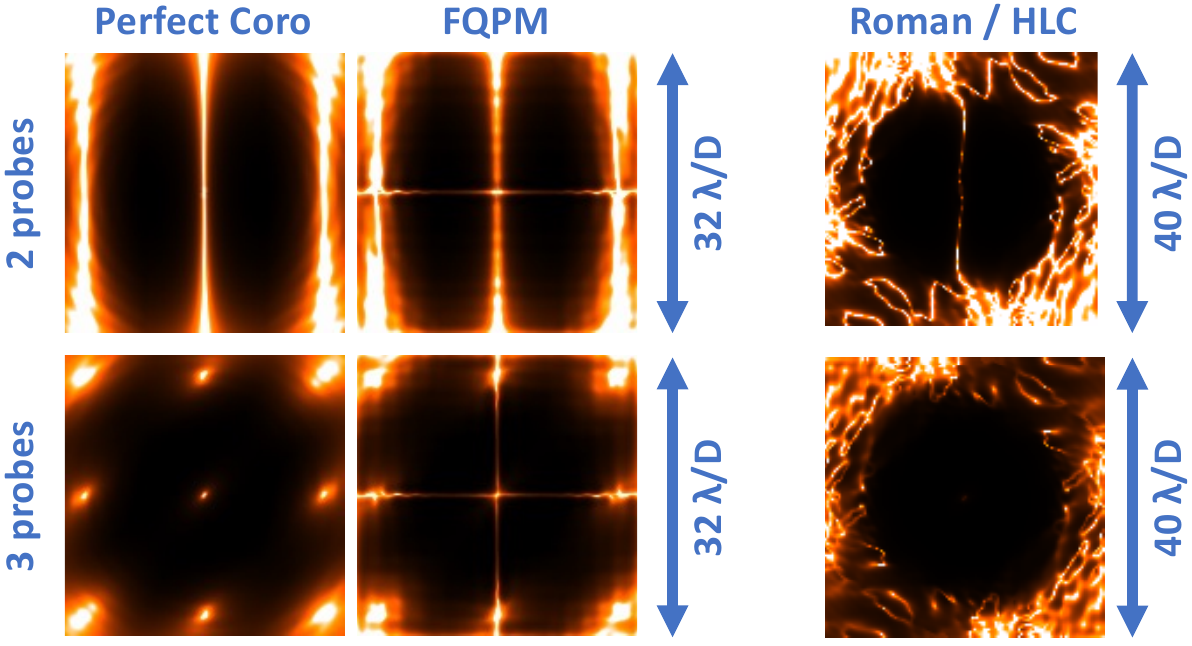}
    \caption{\it Map of the inverse of the minimum eigenvalue for the PWP matrix for each pixel in the focal plane, for 2 or 3 probes and for different coronagraph. The PWP matrix is computed by simulating the effect of pushing neighbor actuators (on a vertical line for 2 and in triangle for 3). Dark zone indicate high eigenvalues : pixels that are well modulated by the method. Bright zone (low eigenvalues) : pixels that are not well modulated. Eigenvalues are dependent on the coronagraph type, color scale is arbitrarily set to show the comparison of the different modulated zones in each case.}
    \label{fig:eigen_pwp} 

\end{figure}

First described by Give’on et al. (2007) \cite{giveon_closed_2007}, PWP applies pairs of known phase perturbations (``probes'') with opposite signs on a deformable mirror, producing intensity modulations in the science image. By comparing the images obtained with the positive and negative probes, the algorithm isolates the interference term between the unknown speckle field and the known probe field, allowing reconstruction of the speckle complex electric field in the focal plane. A detailed description of the PWP formalism was provided by Potier et al. (2020) \cite{potier_increasing_2020}. To recover the electric field from the probed images, a PWP matrix is built using a linearized coronagraphic model, linking, for each pixel in the detector, the difference in probed image intensities to the real and imaginary parts of the complex electric field. For a given pixel, this linear system of equations can only be solved if the pixel is modulated independently by each probe. For this reason, we often calculate the minimum eigenvalue of the PWP matrix for this pixel, where high eigenvalues indicate that the pixel is well-modulated, and low eigenvalues indicate the system is not independent (either a pixel is not modulated by one of the probes or both probes modulate it similarly). This minimum eigenvalue is shown in Fig~\ref{fig:eigen_pwp} for different coronagraph and for different sets of probes (2 or 3 pairs of probed images). Dark region of the image indicate where the speckles are well modulated by the probes and bright regions where the speckles are not well modulated. 

The PWP method is based on a linearized model, and for this reason heavily limited by known non-linearities \cite{laginja_extended_2025} or by unknown error in the underlying model \cite{zhou_roman_2020}. In this new method, we use the same probes to modulate the speckles. Although this new approach does not rely on a linearized model, we use the PWP formalism to ensure that each pixel is modulated "enough" if we use 3 probes so that the nature of each spot (speckle or planet) can be inferred. For this reason, we mostly use the same number probes (3 probes) and only attempt detection in regions well-modulated by the probes. 

\subsection{Data set simulation}

The \texttt{Asterix}\cite{mazoyer_AsterixSimulator} package is a Python tool designed to simulate high-contrast imaging instruments and laboratory testbeds, with a particular emphasis on focal-plane wavefront sensing and wavefront control techniques used in exoplanet imaging.We used \texttt{Asterix} to simulate PWP sequences, comprising 3 pairs of probes and an unprobed focal-plane image. We set the wavelength $\lambda = 575$ nm, and produced purely monochromatic images for this preliminary study. In this proof-of-concept proceeding, we do not introduce photon or detector noise into the images. To produce a PWP sequence, we push a probe, acquire a focal-plane image, pull the probe, acquire a second image, and repeat this for each of the 3 probes. For most of this study, the probes are pushed by 34 nm (PV), in a range where the PWP algorithm should remain linear ($\sim \lambda / 17$). A simulated PWP sequence thus consists of 7 square images, each of size $N_{FP} \times N_{FP}$: 3 pairs of probes and an unprobed image ($N_{FP}$ is the number of pixel in the focal plane). All focal plane intensity images are normalized to the maximum of the off axis stellar PSF. We then randomly inject 0, 1, or 2 planets as off-axis PSFs into each image sequence. Planets are injected at random positions and flux values, with ranges presented in Table~\ref{tab:injection_regions}. We only inject planets in detectable positions, i.e. randomly in the region where the speckles are modulated by the probes (see Fig~\ref{fig:eigen_pwp}). Finally, we produced an associated truth map, a $N_{FP} \times N_{FP}$ map where all pixels are set to zero, except in the injected planet PSFs' core, where they are set to one. This truth map is used to train the algorithm to recognize planets. We simulated several coronagraphs:  

\noindent \textbf{Perfect and FQPM coronagraphs --} For the perfect coronagraph with a clear aperture, only the diffractive effects of the telescope (unaberrated PSF) are subtracted. The Four Quadrant Phase Mask (FQPM\cite{rouan_fqpm_2000}) is also simulated with a clear aperture. We used a sampling of 5 pixels per $\lambda/D$ and focal plane image of $32\lambda/D \times 32\lambda/D$ ($N_{FP} =160$ pixels), to reproduce the sampling on Paris Observatory THD2 experimental platform \cite{Laginja2026THD2}. We also simulated aberrations similar to the ones measured on THD2 (randomly drawn phase aberrations of 15 nm RMS and realistic amplitude aberrations). We use single actuator probes, adjacent and in a triangle shape, close the center of the pupil \cite{potier_increasing_2020}.

\noindent \textbf{Roman's Hybrid Lyot Coronagraph (HLC\cite{trauger_hybrid_2016}) -- }  We used a sampling of 3 pixels per $\lambda/D$ and focal plane image of $40\lambda/D \times 40\lambda/D$ ($N_{FP} =120$ pix), closer the actual sampling of the Roman coronagraphic instrument in Band 1. With flat shapes on the DMs, the HLC contrast is limited by diffraction effects from the aperture. For this reason, we first created a dark hole at a normalized intensity level of $10^{-9}$ between 3 and 9 $\lambda/D$ (Roman HLC target contrast) and then added randomly drawn 5 nm RMS phase aaberrations,which raised the normalized intensity to $10^{-6}$ in the 3-9 $\lambda/D$ region. To modulate speckles, we used Gaussian probes, which have been shown \cite{laginja_extended_2025, delaye_romanprobes2026} to rely on a slightly more linear model than Roman's original sinc-sinc-sin probes. These probes are centered on adjacent actuators in a triangle shape, close to the central obscuration of the Lyot stop and they well-modulate the dark-hole region (Fig~\ref{fig:eigen_pwp}, right). 

\begin{table}[ht]
\centering
\renewcommand{\arraystretch}{1.1}
\begin{tabular}{|l|c|c|c|}
\hline
\textbf{Coronagraph} & \textbf{$\lambda/D$ in pixels} & \textbf{Planet Injection Region ($\lambda/D$)} & \textbf{Contrast Range} \\ \hline
Perfect & 5 & $\sqrt{x^2+y^2} > 1.22 \quad \text{and} \quad \max(|x|, |y|) \leq 10$ & $[5 \times 10^{-5}, 10^{-3}]$ \\ \hline
FQPM    & 5 & $\min(|x|, |y|) > 1.22 \quad \text{and} \quad \max(|x|, |y|) \leq 10$ & $[5 \times 10^{-5}, 10^{-3}]$ \\ \hline
Roman HLC  & 3 & $3 < \sqrt{x^2+y^2} < 9$ & $[2 \times 10^{-6}, 5 \times 10^{-5}]$ \\ \hline
\end{tabular}
\vspace{0.1cm}
\caption{\it Planet injection regions and contrast ranges. Coordinates $(x,y)$ are expressed in units of $\lambda/D$.}
\label{tab:injection_regions}
\end{table}

For each coronagraph, we simulated a total of 2250 ``PWP scenes'' of size $N_{FP} \times N_{FP} \times 7$. This 7-slice structure comprises the detector image and two images (push and pull) for each of the three probes. On average, each scene contains one injected planet embedded among a few hundreds speckles. The dataset is split into 1200 scenes for training and 300 for validation. The remaining 750 scenes were generated using a distinct random seed to serve as a strictly independent test set for algorithm evaluation.

\subsection{EPICX (Exoplanet Patch-based Image Classification with \texttt{XGBoost}) pipeline.}
\begin{figure}[!ht]
    \centering
    \includegraphics[width=\textwidth]{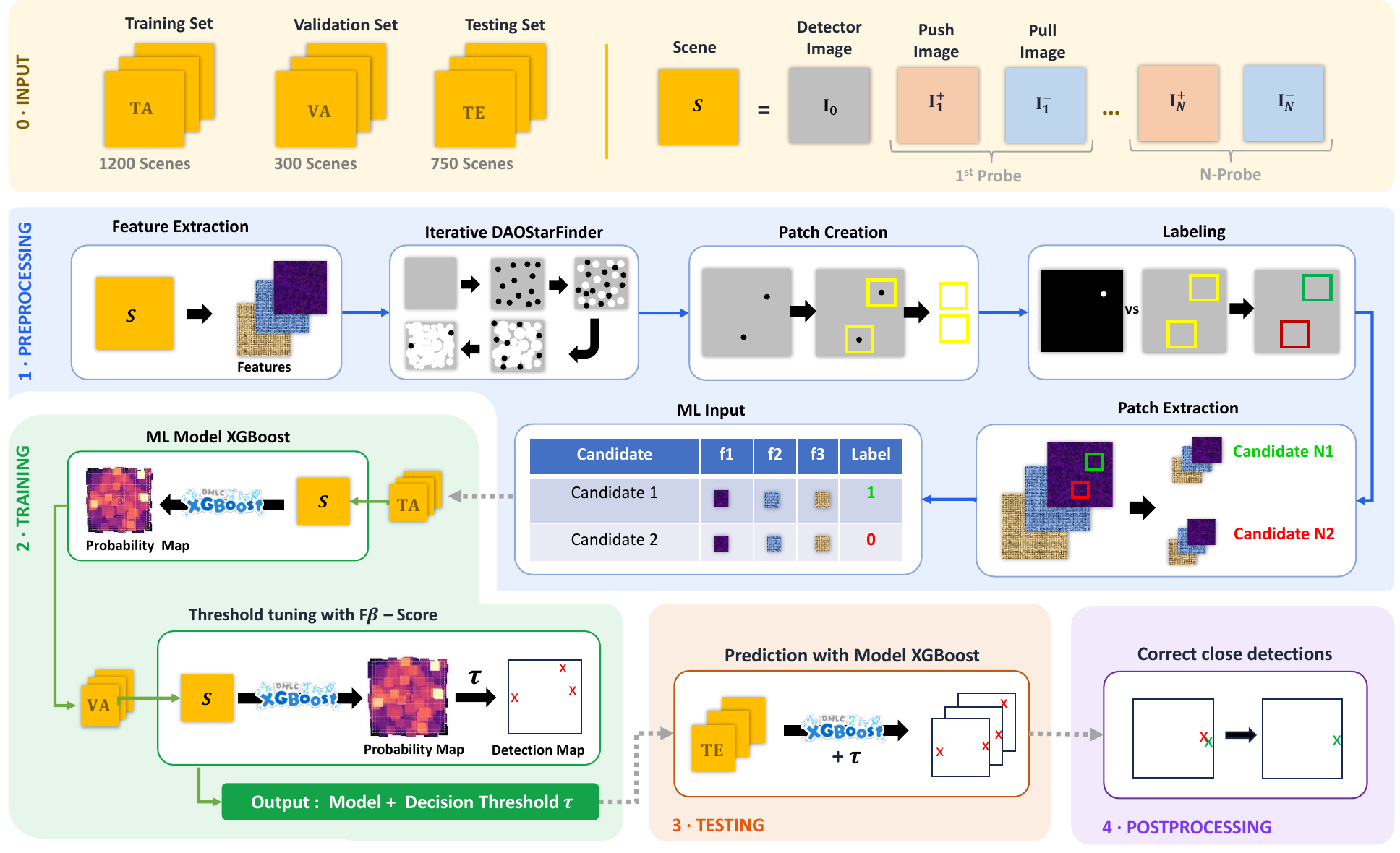}
    \caption{\it \textbf{Schematic overview of EPICX pipeline.} 
    }
    \label{fig:diagram} 
\end{figure}

Fig.~\ref{fig:diagram} shows the detection process for the EPICX algorithm, from identification of PSF-shaped structures in the image to their binary classification as planet or speckle.

\noindent \textbf{\texttt{DAOStarFinder} --} This algorithm is a source-detection tool commonly used in astronomy to identify point-like objects. It implements the DAOFIND algorithm \cite{stetson_daophot_1987}, which is available in the \texttt{Photutils} Python package \cite{bradley_photutils_2026}. The method convolves an image with a Gaussian kernel matching the expected point-spread function (PSF) of stars and then searches for local maxima above a specified detection threshold. The goal is to identify PSF-shaped structures in the images (planets or speckles) that we can feed to the decision tree instead of trying to classify independently each pixel of the image. In this step of the algorithm, we do not apply any selection, we classify all the image illuminated pixels into specific structures of interest. 

\begin{figure}[!ht]
    \centering
    \includegraphics[width=\textwidth]{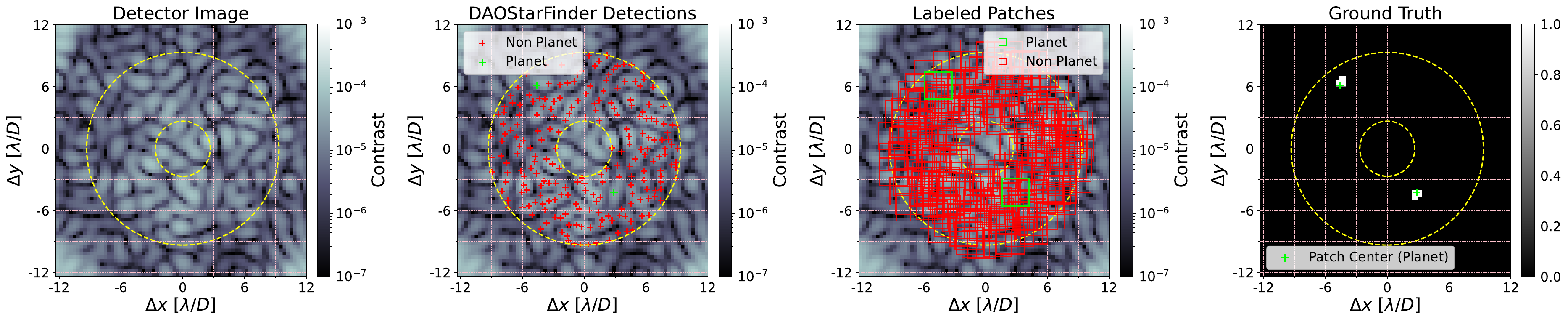}
    \caption{\textbf{Speckle detection and patch labeling process.} \textbf{From left to Right:} Unprobed detector image displayed in logarithmic contrast. Candidate coordinates identified by the iterative \texttt{DAOStarFinder} algorithm, mapping both the injected planet and background speckles. The $N_P \times N_P$ patches defined around each speckle (red patches) or planet (green patches) detected center prior to extraction. The binary ground truth mask indicating the exact location of the injected planet.}
    \label{fig:dao_centers_and_patches} 

\end{figure}

\noindent \texttt{DAOStarFinder} returns a list of coordinates indicating the centers of the PSF-shaped structures in the non-probed image (see Fig.~
\ref{fig:dao_centers_and_patches}). We then subtract a Gaussian centered at these positions and reapply the algorithm to detect PSF-shaped structures in the residuals. We recursively apply \texttt{DAOStarFinder} 4 times, updating the list of PSF-shaped structure centers accordingly. This recursive application ensures that all PSF-shaped structures in the images are included, even in the probable case where planets and speckles overlap. To prevent redundant extraction of the same structure across iterations, we enforce a continuous proximity check: any newly detected source within a minimum separation distance (set to $0.5 \times \textit{FWHM}$) of a previously accumulated center is considered a duplicate and discarded. Finally, after all iterations are complete, we apply a spatial mask to exclude coordinates outside the admissible planet injection regions specified in Table~\ref{tab:injection_regions}, plus a margin of 1 pixel.

\noindent We configure \texttt{DAOStarFinder} using three primary parameters: the \textit{FWHM}, which defines both the initial convolution kernel and the standard deviation of the subtracted 2D Gaussians ; an iterative sequence of \textit{threshold} values ([0, 1e-6, 1e-6, 1e-5]) to progressively detect fainter structures ; and a strict \textit{min\_separation} criterion to prevent overlapping duplicate detections. Additionally, we disable default morphological filters (e.g. roundness and sharpness) to preserve all potential candidates before the machine learning classification stage.

\noindent \textbf{Feature Engineering --} Instead of feeding raw intensity pixels to the decision tree, we first transform the optical information into physically meaningful metrics, known as features. Crucially, this computation is performed prior to patch extraction. The feature extractor processes the entire $N \times N$ field of view using the complete scene, which comprises the unprobed detector image and the stack of probed images. We compute global temporal statistics across the channels and apply a sliding spatial window (kernel) of size $N_k \times N_k$ across the full $N \times N$ images to estimate local properties. 

The kernel size $N_k$ is defined as the nearest higher odd integer that encompasses the expected full core of the Point Spread Function ($\sim 2.44\lambda/D$ in diameter). Using an odd integer ensures the sliding window has a single, unambiguously defined central pixel. This is intentionally larger than the expected PSF core to provide the algorithm with spatial context around the planet/speckle.

To optimize efficiency, an Iterative Feature Elimination (IFE) process reduced our initial pool of 17 features to the 12 most impactful. All retained features detailed in Appendix Table~\ref{tab:selected_features}. As expected from the principles of optical coherence, the most critical features capture probe modulations more than spatial variation: coherent speckles strongly fluctuate as we probe the images, whereas the incoherent planetary signal remains unaffected. This step transforms the raw image inputs into a comprehensive $N \times N \times 12$ feature cube.

\vspace{0.1cm}
\noindent \textbf{Feature Patch Extraction and Labeling --} Once the full-frame 3D feature cube is generated, we extract training samples using the spatial coordinates identified by the iterative \texttt{DAOStarFinder} detection step. For each detected coordinate, we define a bounding box of size $N_P \times N_P$. To maintain perfect spatial consistency with the mathematical transformations applied in the previous step, the patch size is set to equal the spatial kernel size ($N_P = N_k$). 

We apply these bounding boxes to extract the $N_P \times N_P$ patches directly from the 12-layer feature cube. This ensures that every pixel within the extracted candidate patch contains contextual, pre-calculated physical information about its surrounding environment, yielding a tensor of shape $N_P \times N_P \times 12$.

Finally, each feature patch is assigned a binary label by comparing its center to the ground truth map. A candidate is labeled as a planet (1) if its distance to an injected target is $\leq 0.5 \times \textit{FWHM}$; otherwise, it is classified as not a planet (0). To prevent over-representing a single source, we enforce a strict one-to-one mapping: if multiple patches match the same planet, only the closest is retained as a positive match, while redundancies are discarded. Lastly, if a planet is missed during the initial extraction, the nearest non planet candidate is relabeled as a target. This ensures all injected planets are represented without artificially altering the coordinates originally detected by the algorithm.

\noindent \textbf{\texttt{XGBoost}\cite{chen2016xgboost} (eXtreme Gradient Boosting) --} This machine learning algorithm builds an ensemble of decision trees. Instead of creating one large tree, it grows many small trees sequentially. Each new tree is trained to correct the errors made by the previous trees. The predictions from all the trees are then combined, resulting in a model that is often more accurate and robust than a single decision tree, especially for highly imbalanced data sets (speckles far outnumber planets).

\noindent \textbf{Evaluation metrics --} With only a few detections out of the hundreds of stars observed, direct imaging yields particularly low detection rates. We also note that candidates need to be individually reobserved using long sequences on the world's largest telescopes (e.g. VLT, JWST) to be confirmed. Consequently, our algorithm must strictly prioritize minimizing the False Positive Rate (speckles mistaken as planets) to ensure that only the most likely candidates are reobserved. To achieve this, we evaluate the model using metrics specifically suited for highly imbalanced datasets and precision-critical objectives. Since our algorithm outputs a continuous probability map, its performance can be analyzed either at a specific decision threshold, or globally by aggregating performance across all possible thresholds.

\begin{itemize}
    \item \textbf{Evaluation at a specific decision threshold} \texttt{XGBoost} outputs a continuous probability score for each patch. To convert these into binary predictions (planet or speckle), a specific decision threshold must be selected. At any given threshold, we derive the following metrics:
    \begin{itemize}
        \item \textbf{Precision:} The fraction of the algorithm's positive detections that are actual planets. Maximizing precision is our primary goal, as it directly reflects a low false positive rate.
        \item \textbf{Recall:} The fraction of all true injected planets that are successfully detected by the algorithm.
        \item \textbf{$F_\beta$ Score:} A metric that combines both Precision and Recall into a single value. Given that false positives are significantly more detrimental than false negatives (due to the high cost of follow-up observations), we use the $F_\beta$ score with $\beta = 0.5$:
        $$ F_\beta = (1 + \beta^2) \frac{\text{Precision} \cdot \text{Recall}}{(\beta^2 \cdot \text{Precision}) + \text{Recall}} $$
        By setting $\beta = 0.5$, we weight Precision twice as heavily as Recall, mathematically forcing the metric to prioritize a low false positive rate.

    See Appendix \ref{app:metrics} for a more detailed explanation of evaluation metrics.
    \end{itemize}
    
    \item \textbf{Evaluation across all thresholds:} Having defined Precision and Recall for a single threshold, we also evaluate the model's overall capability independently of any specific cut-off. Because speckles overwhelmingly outnumber planets in our dataset, traditional ROC curves are skewed by the massive number of true negatives. We instead use the \textbf{Precision-Recall Area Under the Curve (PR AUC)}, which integrates the Precision-Recall trade-off across the entire range of possible confidence thresholds, providing a single global score of the model's performance.
    \end{itemize}

\noindent Finally, it is important to mention that the decision of this threshold is via a dynamic threshold tuning process during the validation phase, where we select the specific cut-off value that strictly maximizes the $F_{0.5}$ score. This ensures that our final binary classifications align perfectly with our strict observational constraints.

\noindent \textbf{Post-processing --} Since the decision tree evaluates patches independently and patches largely overlap in the image, a single planet may trigger multiple neighboring detections. To consolidate these, we define each detection by its patch center. After applying the decision threshold, we sort all surviving detections by their probability. For any pair of detections, if their centers are separated by a distance of less than $|N_P/2 - 0.5 \times \textit{FWHM}|$, we consider them redundant. We then discard the lower-probability candidate, effectively merging the cluster into a single, unambiguous source centered at the most probable coordinate.

\section{Results}

\begin{figure}[!ht]
    \centering
    \includegraphics[width=0.96\textwidth]{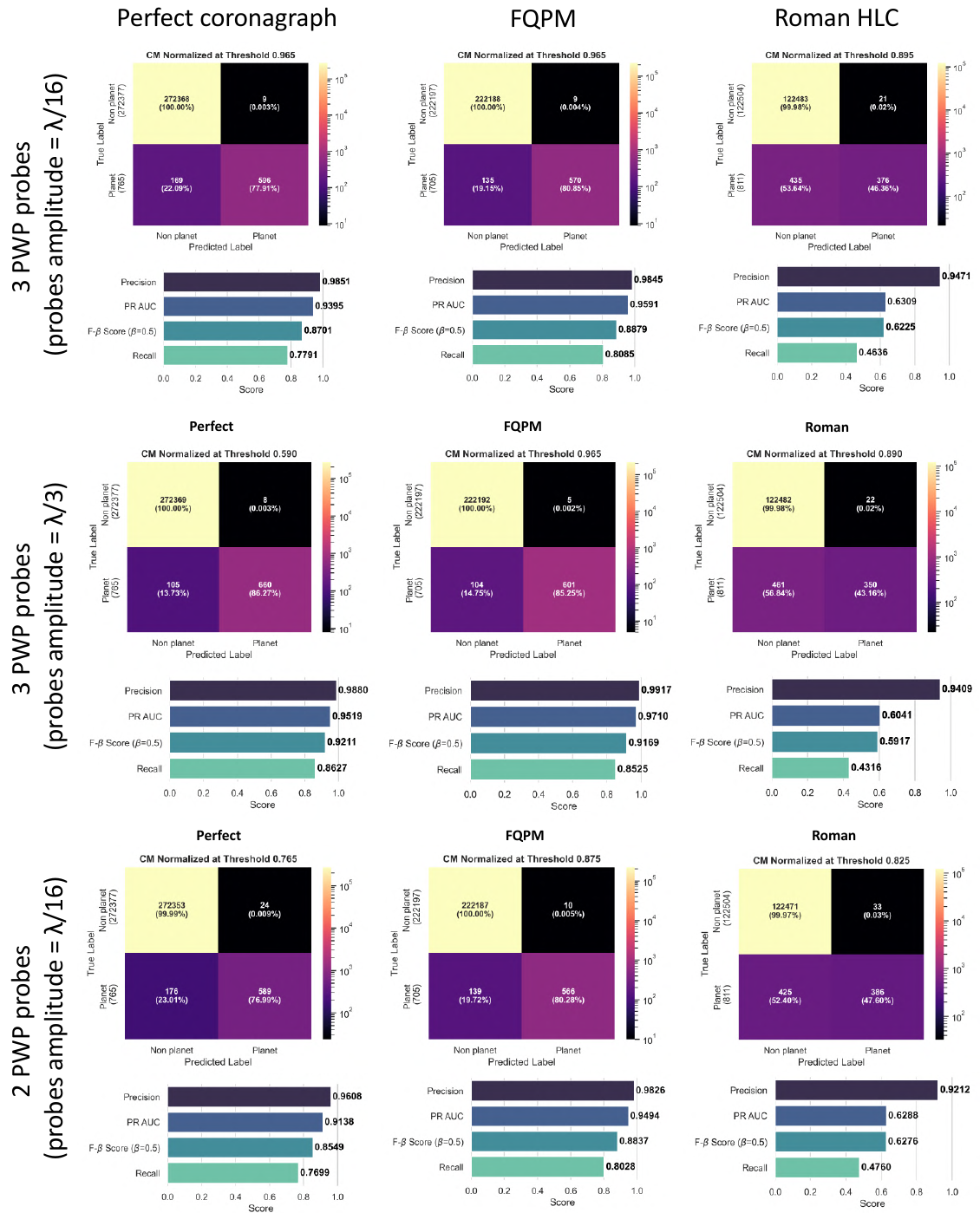}
    \caption{\textbf{Confusion matrices describing performance of the \texttt{XGBoost} pipeline.} Columns represent the simulated coronagraph (Perfect, FQPM, Roman HLC), while rows denote the PWP probe configurations (3 probes with low amplitude, 3 probes with high amplitude, and 2 probes with low amplitude).}
    \label{fig:cm} 
\end{figure}

In this section, we show some preliminary results obtained with the EPICX classification algorithm. To measure the performance of this algorithm, we display the confusion matrix for each experiment. Fig.~\ref{fig:cm} (first row) shows the performance for each coronagraph in the set-up originally described (3 PWP probes each pushed at 36 nm or $\lambda/16$). Each panel displays the testing set confusion matrix at the dynamically tuned threshold that maximizes the $F_{0.5}$ score, alongside its corresponding metrics (Precision, PR AUC, $F_{0.5}$, Recall). Top left quadrant shows the speckles detected as speckles, and bottom right, the planets detected as planets over the testing sample. The other diagonal shows the False Negatives on bottom left (the missed planets) and the False Positives on top right (speckle interpreted as planets). The heavily populated True Negative quadrants highlight the extreme class imbalance (there are many more speckles than planets in the dataset). These matrices show that our goal of minimizing the number of False Positives (i.e., detecting planets where there are only speckles) has been achieved. We note that many planets have been missed ($\sim$ 20\% of False Negatives), but we did not properly quantify the difficulty of each injected planet in this initial tests, as we broadly injected planets at the same level of the speckles. In subsequent tests, we will add realistic photon noise and quantify more precisely the injection brightness in terms of their ratio to the shot noise level.

\newpage

\newpage
\begin{figure}[!ht]
    \centering
    \includegraphics[width=0.96\textwidth]{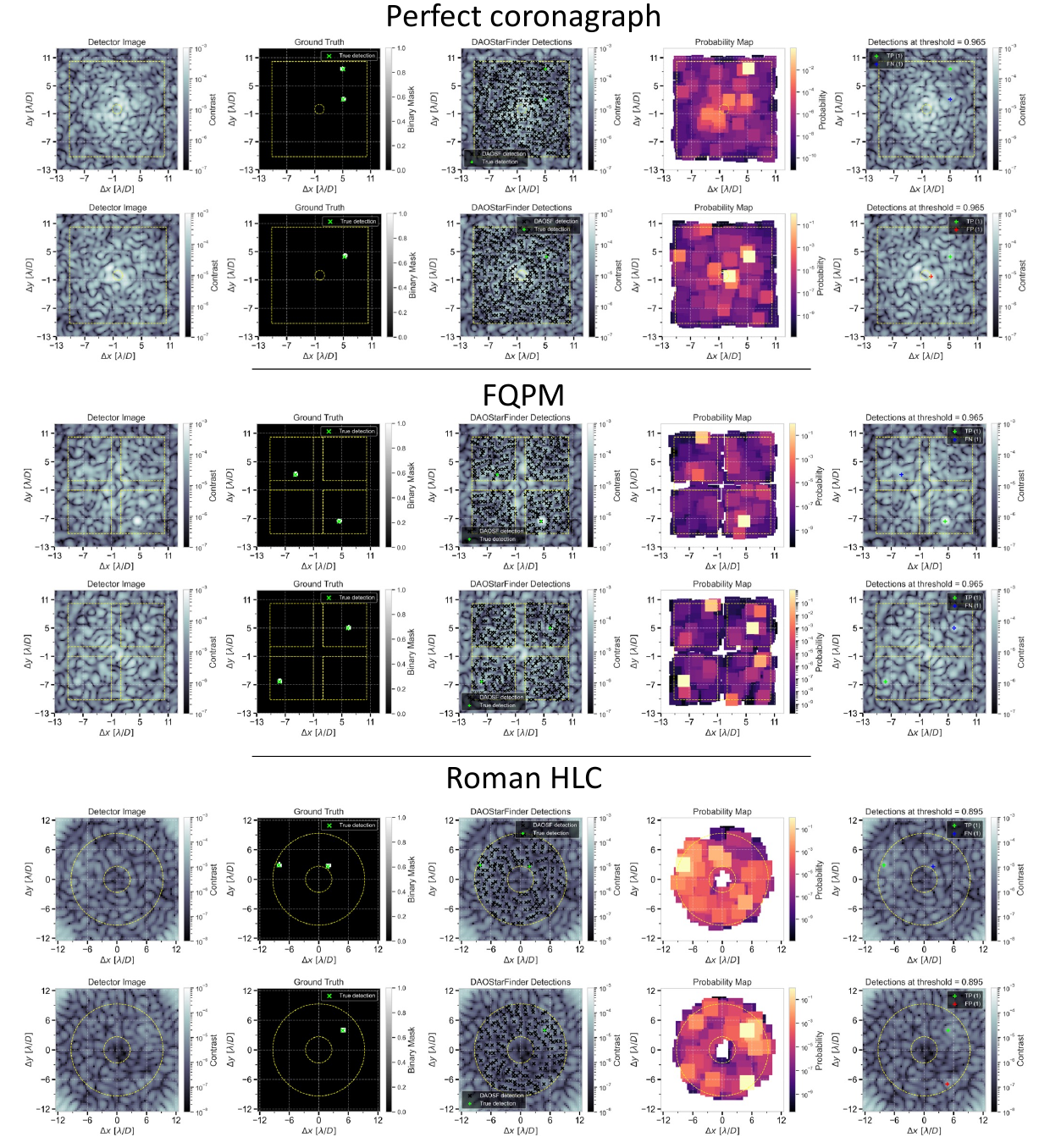}
    \caption{\it Some results in the testing sets obtained for each coronagraph with 3 PWP probes, each pushed with an amplitude of $\lambda/16$. From left to right: original science image, truth map, result of the iterative \texttt{DAOStarFinder}, \texttt{XGBoost} probability map and finally original scene showing detected planet (in green), False Positives (in red) and undetected planet (False Negatives, in blue).}
    \label{fig:epicx_cases} 
\end{figure}

Fig.~\ref{fig:epicx_cases} shows some of the results for each coronagraph, where we detected planets, missed some, or detected false planets. This figure intentionally shows difficult cases to illustrate the limit of EPICX (2 False Negatives in 6 shown cases). In reality, False Negatives are quite rare ($sim 10$ out of 750 PWP scenes in the testing set, see Fig.~\ref{fig:cm}. We draw attention to the fact that the injected planets are completely hidden in the original speckled images and are detectable only using PWP scenes. The comparison of this algorithm to the classical linear PWP-CDI will be performed in a subsequent study. 

However, we performed some tests showing exciting prospects for this algorithm. To test the (non-)linearity of this approach, we performed the analysis by pushing the PWP not by 36 nm (or $\lambda/16$), but by 192 nm (or $\lambda/3$) well above the range normally accepted by classical PWP-CDI, which relies on the small phase hypothesis to measure a linear model. The results are shown in the second row of Fig.~\ref{fig:cm} and are completely similar to those with a small PWP amplitude. As we currently simulated this problem without photon noise, the advantage of pushing the probes to high amplitudes is not immediately obvious. However, in a real-case scenario, modulating the speckles more means we can perform PWP much more effectively: we can either measure the speckles more accurately with the same exposure time or achieve the same measurement with shorter probes \cite{delaye_romanprobes2026}).

Finally, we performed a test where we used only 2 PWP probes showing a small degradation in performance. This is expected as, in this case, some of the speckles are no longer modulated (see Fig.~\ref{fig:eigen_pwp}). In this case, we only acquire 5 images (the original scene and 2 pairs of probes) instead of 7. This is likely interesting in cases where we possess some information on the system orbits, we could apply a shorter CDI to only part of the system in the image, saving telescope time. 

 

\section{Conclusions}

The EPICX algorithm presented in this proceeding demonstrates that machine learning trained on a relatively small dataset can effectively learn the modulation model and separate planets from speckles in PWP detection. This is a particularly interesting to allow faster speckles calibration compared to existing ADI (in ground) and roll + RDI (in space) image. Finally, compared to classical CDI algorithms\cite{potier_coherent_2025}, it has the potential to minimize model errors and allows escaping linear constraints.

The next steps are to gradually develop a more complex simulation model by adding photon noise and training on more realistic Roman or SPHERE+ simulators, in order to precisely quantify the gain of this method compared to existing algorithms currently used on coronagraphic instruments. In parallel, we will demonstrate its performance on an experimental platform (the THD2 experimental testbed \cite{Laginja2026THD2}), opening the path to an on-sky validation on telescopes with PWP scenes (e.g. Roman/CGI or VLT/SPHERE+).

\acknowledgments 
 
This project has received funding from the European Research Council (ERC) under the European Union's Horizon Europe research and innovation programme under the grant agreement \#101230218. M.J.M.T. acknowledges support from the visiting program of the CNRS French-Chilean Laboratory for Astrophysics (FCLA, IRL-3386) for providing a travel grant to work in France.

\noindent \textit{Software} - This study made use of the following Python packages: \texttt{Asterix}\cite{mazoyer_AsterixSimulator}, 
\texttt{Astropy}\cite{astropy_collaboration_2022}, 
\texttt{Matplotlib} \cite{hunter2007matplotlib}, \texttt{NumPy}\cite{harris2020NumPy}, 
\texttt{Photutils} \cite{bradley_photutils_2026},  \texttt{SciPy}\cite{virtanen2020scipy},
\texttt{SHAP}\cite{lundberg2017shap},
\texttt{XGBoost}\cite{chen2016xgboost}. 
We also acknowledge the use of Gemini and MistralAI to enhance the python code presented in this proceeding.

\bibliography{report} 

@article{soummer_detection_2012,
	title = {Detection and Characterization of Exoplanets and Disks Using Projections on Karhunen-Loève Eigenimages},
	volume = {755},
	issn = {2041-8205},
	url = {http://iopscience.iop.org/2041-8205/755/2/L28},
	doi = {10.1088/2041-8205/755/2/L28},
	pages = {L28},
	number = {2},
	journal = {The Astrophysical Journal Letters},
	shortjournal = {{ApJL}},
	author = {Soummer, Rémi and Pueyo, Laurent and Larkin, James},
	urldate = {2014-04-07},
	date = {2012-08-20},
	langid = {english},
  year = {2012},
  month = {aug},
}

@article{nishikawa_coherent_2022,
	title = {The coherent differential imaging on speckle area nulling ({CDI}-{SAN}) method for high-contrast imaging under speckle variation},
	volume = {930},
	issn = {0004-637X, 1538-4357},
	url = {http://arxiv.org/abs/2204.10002},
	doi = {10.3847/1538-4357/ac5f44},
	pages = {163},
	number = {2},
	journal = {The Astrophysical Journal},
	shortjournal = {{ApJ}},
	author = {Nishikawa, Jun},
	urldate = {2022-08-12},
	date = {2022-05-01},
	eprinttype = {arxiv},
	eprint = {2204.10002 [astro-ph]},
  year = {2022},
  month = {may},
}

@ARTICLE{Cantero_NA_SODINN_2023,
       author = {{Cantero}, C. and {Absil}, O. and {Dahlqvist}, C.-H. and {Van Droogenbroeck}, M.},
        title = "{NA-SODINN: A deep learning algorithm for exoplanet image detection based on residual noise regimes}",
      journal = {Astronomy and Astrophysics},
         year = 2023,
        month = dec,
       volume = {680},
          eid = {A86},
        pages = {A86},
          doi = {10.1051/0004-6361/202346085},
archivePrefix = {arXiv},
       eprint = {2302.02854},
 primaryClass = {astro-ph.IM},
       adsurl = {https://ui.adsabs.harvard.edu/abs/2023A&A...680A..86C}
}

@article{potier_increasing_2020,
	title = {Increasing the raw contrast of {VLT}/{SPHERE} with the dark hole technique. I. Simulations and validation on the internal source},
	volume = {638},
	issn = {0004-6361},
	url = {http://adsabs.harvard.edu/abs/2020A%26A...638A.117P},
	doi = {10.1051/0004-6361/202038010},
	pages = {A117},
	journal = {Astronomy and Astrophysics},
	shortjournal = {A\&A},
	author = {Potier, A. and Galicher, R. and Baudoz, P. and Huby, E. and Milli, J. and Wahhaj, Z. and Boccaletti, A. and Vigan, A. and N'Diaye, M. and Sauvage, J.-F.},
	urldate = {2020-09-14},
	date = {2020-06-01},
	year = {2020},
	month = {jun},
}

@article{poberezhskiy_romancgi_2025,
	title = {Overview of Roman Coronagraph Instrument requirements, test campaign, and results},
	volume = {11},
	url = {https://ui.adsabs.harvard.edu/abs/2025JATIS..11c1511P},
	doi = {10.1117/1.JATIS.11.3.031511},
	pages = {031511},
	journal = {Journal of Astronomical Telescopes, Instruments, and Systems},
	author = {Poberezhskiy, Ilya and Cady, Eric and Heydorff, Kathryn and Kern, Brian and Luchik, Thomas and Zhao, Feng and Bailey, Vanessa and Bush, Nathan and Colavita, Mark and Creager, Brandon and Fathpour, Nanaz and Gaidon, Clement and Grue, Amanda and Kempenaar, Joshua and Krist, John and Kuan, Gary and Lam, Jonathan and Mandić, Milan and Marx, David and Nemati, Bijan and Eldorado Riggs, A. J. and Seo, Byoung-Joon and Shi, Fang and Smith, Matthew W. and Zhou, Hanying},
	urldate = {2026-06-22},
	date = {2025-07-01},
	year = {2025},
	month = {7}
}

@article{gerard_cdiscc_2018,
	title = {Fast Coherent Differential Imaging on Ground-based Telescopes Using the Self-coherent Camera},
	volume = {156},
	url = {https://ui.adsabs.harvard.edu/abs/2018AJ....156..106G/abstract},
	doi = {10.3847/1538-3881/aad23e},
	pages = {106},
	number = {3},
	journal = {The Astronomical Journal},
	shortjournal = {{AJ}},
	author = {Gerard, Benjamin L. and Marois, Christian and Galicher, Raphael},
	urldate = {2018-08-31},
	date = {2018-09},
	year = {2018},
	month = {9},
	langid = {english},
}

@article{cady_romanhowfsc_2025,
	title = {High-order wavefront sensing and control for the Roman Coronagraph Instrument ({CGI}): architecture and measured performance},
	volume = {11},
	url = {https://ui.adsabs.harvard.edu/abs/2025JATIS..11b1408C},
	doi = {10.1117/1.JATIS.11.2.021408},
	shorttitle = {High-order wavefront sensing and control for the Roman Coronagraph Instrument ({CGI})},
	pages = {021408},
	journal = {Journal of Astronomical Telescopes, Instruments, and Systems},
	author = {Cady, Eric and Bowman, Nicholas and Greenbaum, Alexandra Z. and Ingalls, James G. and Kern, Brian and Krist, John and Marx, David and Poberezhskiy, Ilya and Eldorado Riggs, A. J. and Ruane, Garreth and Seo, Byoung-Joon and Shi, Fang and Zhou, Hanying},
	urldate = {2026-06-22},
	date = {2025-04-01},
	year = {2025},
	month = {4}
}

@online{mazoyer_AsterixSimulator,
	title = {Asterix: Simulate your High-Contrast Instruments},
	url = {https://asterix-hci.readthedocs.io/},
	author = {Mazoyer, J. and Potier, A. and Laginja, I. and Galicher, R.},
	urldate = {2024-01-05},
	year = {2019},
    note = {\url{https://asterix-hci.readthedocs.io/}},
}

@article{astropy_collaboration_2022,
    title = {The Astropy Project: Sustaining and Growing a Community-oriented Open-source Project and the Latest Major Release (v5.0) of the Core Package},
	volume = {935},
	issn = {0004-637X},
	url = {https://ui.adsabs.harvard.edu/abs/2022ApJ...935..167A},
	doi = {10.3847/1538-4357/ac7c74},
	shorttitle = {The Astropy Project},
	pages = {167},
	journal = {The Astrophysical Journal},
	publisher = {{IOP}},
	author = {{Astropy Collaboration et al.}},	
	urldate = {2026-06-22},
	date = {2022-08-01},
	year = {2022},
	month = {08},
}

@article{giveon_closed_2007,
	title = {Closed loop, {DM} diversity-based, wavefront correction algorithm for high contrast imaging systems},
	volume = {15},
	issn = {1094-4087},
	url = {https://ui.adsabs.harvard.edu/abs/2007OExpr..1512338G},
	doi = {10.1364/OE.15.012338},
	pages = {12338},
	journal = {Optics Express},
	author = {Give'on, Amir and Belikov, Ruslan and Shaklan, Stuart and Kasdin, Jeremy},
	urldate = {2025-01-09},
	date = {2007-09-01},
	year = {2007},
	month = {09},
	note = {{ADS} Bibcode: 2007OExpr..1512338G},
}

@ARTICLE{Lafreniere2007_LOCI,
       author = {{Lafreni{\`e}re}, David and {Marois}, Christian and {Doyon}, Ren{\'e} and {Nadeau}, Daniel and {Artigau}, {\'E}tienne},
        title = "{A New Algorithm for Point-Spread Function Subtraction in High-Contrast Imaging: A Demonstration with Angular Differential Imaging}",
      journal = {The Astrophysical Journal},
         year = 2007,
        month = may,
       volume = {660},
       number = {1},
        pages = {770-780},
          doi = {10.1086/513180},
archivePrefix = {arXiv},
       eprint = {astro-ph/0702697},
 primaryClass = {astro-ph},
       adsurl = {https://ui.adsabs.harvard.edu/abs/2007ApJ...660..770L}
}

@ARTICLE{Snellen2015_molecularmapping,
       author = {{Snellen}, I. and {de Kok}, R. and {Birkby}, J.~L. and {Brandl}, B. and {Brogi}, M. and {Keller}, C. and {Kenworthy}, M. and {Schwarz}, H. and {Stuik}, R.},
        title = "{Combining high-dispersion spectroscopy with high contrast imaging: Probing rocky planets around our nearest neighbors}",
      journal = {Astronomy and Astrophysics},
         year = 2015,
        month = apr,
       volume = {576},
          eid = {A59},
        pages = {A59},
          doi = {10.1051/0004-6361/201425018},
archivePrefix = {arXiv},
       eprint = {1503.01136},
 primaryClass = {astro-ph.EP},
       adsurl = {https://ui.adsabs.harvard.edu/abs/2015A&A...576A..59S}
}

@article{flasseur_paco_2020,
	title = {{PACO} {ASDI}: an algorithm for exoplanet detection and characterization in direct imaging with integral field spectrographs},
	volume = {637},
	issn = {0004-6361},
	url = {http://adsabs.harvard.edu/abs/2020A%26A...637A...9F},
	doi = {10.1051/0004-6361/201937239},
	shorttitle = {{PACO} {ASDI}},
	pages = {A9},
	journal = {Astronomy and Astrophysics},
	shortjournal = {Astronomy and Astrophysics},
	author = {Flasseur, Olivier and Denis, Loïc and Thiébaut, Eric and Langlois, Maud},
	urldate = {2020-06-09},
	date = {2020-05-01},
	year = {2020},
	month = {05},
}

@inproceedings{chen2016xgboost,
  title={XGBoost: A Scalable Tree Boosting System},
  author={Chen, Tianqi and Guestrin, Carlos},
  booktitle={Proceedings of the 22nd ACM SIGKDD International Conference on Knowledge Discovery and Data Mining},
  pages={785--794},
  year={2016},
  doi={10.1145/2939672.2939785}
}

@inproceedings{baudoz_laboratory_2013,
	title = {Laboratory tests of planet signal extraction in high contrast images},
	url = {http://adsabs.harvard.edu/abs/2013aoel.confE.109B},
	doi = {10.12839/AO4ELT3.13701},
	eventtitle = {Proceedings of the Third {AO}4ELT Conference},
	pages = {109},
	booktitle = {Proceedings of the Third {AO}4ELT Conference},
	author = {Baudoz, Pierre and Mazoyer, Johan and Galicher, Raphael},
	urldate = {2016-02-17},
	date = {2013-12-01},
  year = {2013},
  month = {dec}
}

@article{marois_angular_2006,
	title = {Angular Differential Imaging: A Powerful High-Contrast Imaging Technique},
	volume = {641},
	issn = {0004-637X},
	url = {http://iopscience.iop.org/0004-637X/641/1/556},
	doi = {10.1086/500401},
	shorttitle = {Angular Differential Imaging},
	pages = {556},
	number = {1},
	journal = {The Astrophysical Journal},
	shortjournal = {{ApJ}},
	author = {Marois, Christian and Lafrenière, David and Doyon, René and Macintosh, Bruce and Nadeau, Daniel},
	urldate = {2014-04-07},
	date = {2006-04-10},
	langid = {english},
  year = {2006},
  month = {apr},
}

@article{potier_increasing_2022,
	title = {Increasing the raw contrast of {VLT}/{SPHERE} with the dark hole technique. {II}. On-sky wavefront correction and coherent differential imaging},
	volume = {665},
	issn = {0004-6361},
	url = {https://ui.adsabs.harvard.edu/abs/2022A&A...665A.136P/abstract},
	doi = {10.1051/0004-6361/202244185},
	pages = {A136},
	journal = {Astronomy and Astrophysics},
	shortjournal = {A\&A},
	author = {Potier, A. and Mazoyer, J. and Wahhaj, Z. and Baudoz, P. and Chauvin, G. and Galicher, R. and Ruane, G.},
	urldate = {2022-10-27},
	date = {2022-09},
	langid = {english},
  year = {2022},
  month = {sep},
}

@article{flasseur_deep_2024,
	title = {deep {PACO}: combining statistical models with deep learning for exoplanet detection and characterization in direct imaging at high contrast},
	volume = {527},
	issn = {0035-8711},
	url = {https://ui.adsabs.harvard.edu/abs/2024MNRAS.527.1534F},
	doi = {10.1093/mnras/stad3143},
	shorttitle = {deep {PACO}},
	pages = {1534--1562},
	journal = {Monthly Notices of the Royal Astronomical Society},
	shortjournal = {{MNRAS}},
	publisher = {{OUP}},
	author = {Flasseur, Olivier and Bodrito, Théo and Mairal, Julien and Ponce, Jean and Langlois, Maud and Lagrange, Anne-Marie},
	urldate = {2024-06-27},
	date = {2024-01-01},
	note = {{ADS} Bibcode: 2024MNRAS.527.1534F},
  year = {2024},
  month = {jan},
}

@article{galicher_imaging_2024,
	title = {Imaging exoplanets with coronagraphic instruments},
	volume = {24},
	issn = {1631-0705},
	url = {https://ui.adsabs.harvard.edu/abs/2024CRPhy..24S.133G},
	doi = {10.5802/crphys.133},
	pages = {133},
    journal = {Comptes Rendus Physique},
    shortjournal = {C. R. Phys.},
	author = {Galicher, Raphaël and Mazoyer, Johan},
	urldate = {2024-06-19},
	date = {2024-01-01},
	note = {{ADS} Bibcode: 2024CRPhy..24S.133G},
  year = {2024},
  month = {jan},
}

@article{amara_pynpoint_2012,
	title = {{PYNPOINT}: an image processing package for finding exoplanets},
	volume = {427},
	issn = {0035-8711},
	url = {http://adsabs.harvard.edu/abs/2012MNRAS.427..948A},
	doi = {10.1111/j.1365-2966.2012.21918.x},
	shorttitle = {{PYNPOINT}},
	pages = {948--955},
	journal = {Monthly Notices of the Royal Astronomical Society},
	shortjournal = {{MNRAS}},
	author = {Amara, Adam and Quanz, Sascha P.},
	urldate = {2020-11-24},
	date = {2012-12-01},
  year = {2012},
  month = {dec},
}

@article{laginja_extended_2025,
	title = {Extended linearity in the high-order wavefront sensor for the Roman Coronagraph},
	volume = {698},
	issn = {0004-6361},
	url = {https://ui.adsabs.harvard.edu/abs/2025A&A...698A.130L},
	doi = {10.1051/0004-6361/202553797},
	pages = {A130},
	journal = {Astronomy and Astrophysics},
	author = {Laginja, Iva and Baudoz, Pierre and Mazoyer, Johan and Potier, Axel and Galicher, Raphaël and Boussaha, Faouzi},
	urldate = {2025-09-12},
	date = {2025-06-01},
	note = {{ADS} Bibcode: 2025A\&A...698A.130L},
  year = {2025},
  month = {jun},
}

@article{potier_coherent_2025,
	title = {Coherent differential imaging of high-contrast extended sources with {VLT}/{SPHERE}},
	volume = {704},
	issn = {0004-6361},
	url = {https://ui.adsabs.harvard.edu/abs/2025A&A...704A..98P},
	doi = {10.1051/0004-6361/202556606},
	pages = {A98},
	journal = {Astronomy and Astrophysics},
	publisher = {{EDP}},
	author = {Potier, Axel and Galicher, Raphaël and Baudoz, Pierre and Mazoyer, Johan and Wahhaj, Zahed and Tandon, Ruben and Kühn, Jonas G. and Perez, Laura and Chauvin, Gael},
	urldate = {2026-04-03},
	date = {2025-12-01},
	note = {{ADS} Bibcode: 2025A\&A...704A..98P},
  year = {2025},
  month = {dec},
}

@ARTICLE{racine1999_SDI,
       author = {{Racine}, Ren{\'e} and {Walker}, Gordon A.~H. and {Nadeau}, Daniel and {Doyon}, Ren{\'e} and {Marois}, Christian},
        title = "{Speckle Noise and the Detection of Faint Companions}",
      journal = {Publications of the Astronomical Society of the Pacific},
         year = 1999,
        month = 05,
       volume = {111},
       number = {759},
        pages = {587-594},
          doi = {10.1086/316367},
       adsurl = {https://ui.adsabs.harvard.edu/abs/1999PASP..111..587R}
}

@ARTICLE{bottom2017_CDI,
       author = {{Bottom}, M. and {Wallace}, J.~K. and {Bartos}, R.~D. and {Shelton}, J.~C. and {Serabyn}, E.},
        title = "{Speckle suppression and companion detection using coherent differential imaging}",
      journal = {Monthly Notices of the Royal Astronomical Society},
         year = 2017,
        month = 01,
       volume = {464},
       number = {3},
        pages = {2937-2951},
          doi = {10.1093/mnras/stw2544},
archivePrefix = {arXiv},
       eprint = {1610.00606},
 primaryClass = {astro-ph.IM},
       adsurl = {https://ui.adsabs.harvard.edu/abs/2017MNRAS.464.2937B}
}

@INPROCEEDINGS{baba2005_PDI,
       author = {{Baba}, Naoshi and {Murakami}, Naoshi and {Tate}, Youko and {Sato}, Yoichiro and {Tamura}, Motohide},
        title = "{Objective spectrometer for exoplanets based on nulling polarization interferometry}",
    booktitle = {Techniques and Instrumentation for Detection of Exoplanets II},
         year = 2005,
       editor = {{Coulter}, Daniel R.},
       series = {Society of Photo-Optical Instrumentation Engineers (SPIE) Conference Series},
       volume = {5905},
        month = 08,
        pages = {347-351},
          doi = {10.1117/12.615623},
       adsurl = {https://ui.adsabs.harvard.edu/abs/2005SPIE.5905..347B}
}

@ARTICLE{wahhaj2021_RDI,
       author = {{Wahhaj}, Z. and {Milli}, J. and {Romero}, C. and {Cieza}, L. and {Zurlo}, A. and {Vigan}, A. and {Pe{\~n}a}, E. and {Valdes}, G. and {Cantalloube}, F. and {Girard}, J. and {Pantoja}, B.},
        title = "{A search for a fifth planet around HR 8799 using the star-hopping RDI technique at VLT/SPHERE}",
      journal = {Astronomy \& Astrophysics},
         year = 2021,
        month = 04,
       volume = {648},
          eid = {A26},
        pages = {A26},
          doi = {10.1051/0004-6361/202038794},
archivePrefix = {arXiv},
       eprint = {2101.08268},
 primaryClass = {astro-ph.EP},
       adsurl = {https://ui.adsabs.harvard.edu/abs/2021A&A...648A..26W}
}

@ARTICLE{choquet2016_RDI,
       author = {{Choquet}, {\'E}lodie and {Perrin}, Marshall D. and {Chen}, Christine H. and {Soummer}, R{\'e}mi and {Pueyo}, Laurent and {Hagan}, James B. and {Gofas-Salas}, Elena and {Rajan}, Abhijith and {Golimowski}, David A. and {Hines}, Dean C. and {Schneider}, Glenn and {Mazoyer}, Johan and {Augereau}, Jean-Charles and {Debes}, John and {Stark}, Christopher C. and {Wolff}, Schuyler and {N'Diaye}, Mamadou and {Hsiao}, Kevin},
        title = "{First Images of Debris Disks around TWA 7, TWA 25, HD 35650, and HD 377}",
      journal = {The Astrophysical Journall},
         year = 2016,
        month = 01,
       volume = {817},
       number = {1},
          eid = {L2},
        pages = {L2},
          doi = {10.3847/2041-8205/817/1/L2},
archivePrefix = {arXiv},
       eprint = {1512.02220},
 primaryClass = {astro-ph.SR},
       adsurl = {https://ui.adsabs.harvard.edu/abs/2016ApJ...817L...2C}
}

@ARTICLE{beuzit1997_RDI,
       author = {{Beuzit}, J. -L. and {Mouillet}, D. and {Lagrange}, A. -M. and {Paufique}, J.},
        title = "{A stellar coronograph for the COME-ON-PLUS adaptive optics system}",
      journal = {Astronomy \& Astrophysicss},
         year = 1997,
        month = 10,
       volume = {125},
        pages = {175-182},
          doi = {10.1051/aas:1997370},
       adsurl = {https://ui.adsabs.harvard.edu/abs/1997A&AS..125..175B}
}

@software{bradley_photutils_2026,
	title = {Photutils},
	url = {https://zenodo.org/records/19636730},
	doi = {10.5281/zenodo.19636730},
	publisher = {Zenodo},
	author = {Bradley, Larry and Sipőcz, Brigitta M. and Robitaille, T. P. and Tollerud, E. J. and Vinícius, Zé and Deil, Christoph and Barbary, Kyle and Wilson, Tom J. and Busko, Ivo and Donath, Axel and Günther, Hans Moritz and Cara, Mihai and Lim, P. L. and Meßlinger, Sebastian and Conseil, Simon and Droettboom, Michael and Bostroem, K. Azalee and Bray, E. M. and Bratholm, Lars Andersen and Burnett, Zach and Jamieson, William and Ginsburg, Adam and Taranu, Dan and Barentsen, Geert and Craig, Matthew W. and Morris, Brett M. and Perrin, Marshall and Rathi, Shivangee},
	urldate = {2026-06-18},
	date = {2026-04-18},
  year = {2026},
  month = {apr},
  note = {{DOI : 10.5281/zenodo.19636730}}
}

@inproceedings{Mazoyer2026SAXOplusNCPA,
author = {Mazoyer, Johan and Goulas, Charles and Galicher, Raphaël and
Vidal, Fabrice and Potier, Axel and Ferreira, Florian and
Sevin, Arnaud and Béchet, Clémentine and Dinis, Isaac and
Boccaletti, Anthony and Chauvin, Ga{"e}l and Cortecchia, Fausto and
Galland, Nicolas and Kulcs{'a}r, Caroline and Langlois, Maud and
Lombini, Matteo and Milli, Julien and N'Diaye, Mamadou and
Raynaud, Henri-Fran{\c c}ois and Schreiber, Laura and
Tallon, Michel and Wildi, Fran{\c c}ois},
eventtitle = {Adaptive Optics Systems X},
title = {SAXO+, the second-stage adaptive optics for SPHERE:
NCPA compensation and dark hole loop with a pyramid wavefront sensor},
booktitle = {Proceedings of SPIE Astronomical Telescopes + Instrumentation},
year = {2026},
publisher = {{SPIE}},
pages = {14150-105},
note = {in this proceedings}
}

@inproceedings{Laginja2026THD2,
author = {Laginja, Iva and  Baudoz, Pierre and Potier, Axel and Mazoyer, Johan  and Galicher, Raphaël  and Por, Emiel H.  and Soummer, Rémi  and Sevin, Arnaud  and Pougheon, Erin  and Paviot, Corentin  and Doelman, David  and Landman, Rico  and Snik, Frans  and Bettonvil, Felix  and Rietjens, Jeroen  and van Dijk, Chris  and Peeters, Kristien  and Eigenraam, Alexander  and Krasteva, Mariya  and Taccola, Matteo  },
eventtitle = {Space Telescopes and Instrumentation 2026: Optical, Infrared, and Millimeter Wave},
title = {First high-contrast results on THD2 testbed after infrastructure upgrade },
booktitle = {Proceedings of SPIE Astronomical Telescopes + Instrumentation},
year = {2026},
publisher = {{SPIE}},
pages = { 14145-70},
}

@inproceedings{delaye_romanprobes2026,
author = {Delaye, Lukas and Laginja, Iva and  Baudoz, Pierre and Potier, Axel and Mazoyer, Johan  and Galicher, Raphaël  and Redmond, Susan F. and Lau, Alexis and Riggs, A J Eldorado  and Sirbu, Dan and Por, Emiel H.  and Soummer, Rémi and Pueyo, Laurent},
eventtitle = {Space Telescopes and Instrumentation 2026: Optical, Infrared, and Millimeter Wave},
title = {Enhanced wavefront sensing for the Roman Coronagraph Instrument: Gaussian probes and compact model validation},
booktitle = {Proceedings of SPIE Astronomical Telescopes + Instrumentation},
year = {2026},
publisher = {{SPIE}},
pages = {14145-181},
}

@article{trauger_hybrid_2016,
	title = {Hybrid Lyot coronagraph for {WFIRST}-{AFTA}: coronagraph design and performance metrics},
	volume = {2},
	url = {http://adsabs.harvard.edu/abs/2016JATIS...2a1013T},
	doi = {10.1117/1.JATIS.2.1.011013},
	shorttitle = {Hybrid Lyot coronagraph for {WFIRST}-{AFTA}},
	pages = {011013},
	journal = {Journal of Astronomical Telescopes, Instruments, and Systems},
	shortjournal = {{JATIS}},
	author = {Trauger, John and Moody, Dwight and Krist, John and Gordon, Brian},
	urldate = {2020-02-17},
	date = {2016-01-01},
	year = {2016},
	month = {01},
}

@article{rouan_fqpm_2000,
	title = {The Four-Quadrant Phase-Mask Coronagraph. I. Principle},
	volume = {112},
	issn = {0004-6280},
	url = {http://adsabs.harvard.edu/abs/2000PASP..112.1479R},
	doi = {10.1086/317707},
	pages = {1479--1486},
	journal = {Publications of the Astronomical Society of the Pacific},
	shortjournal = {{PASP}},
	author = {Rouan, D. and Riaud, P. and Boccaletti, A. and Clénet, Y. and Labeyrie, A.},
	urldate = {2016-02-15},
	date = {2000-11-01},
	year = {2000},
	month = {11},
}

@inproceedings{zhou_roman_2020,
	title = {Roman {CGI} testbed {HOWFSC} modeling and validation},
	volume = {11443},
	url = {https://www.spiedigitallibrary.org/conference-proceedings-of-spie/11443/114431W/Roman-CGI-testbed-HOWFSC-modeling-and-validation/10.1117/12.2561087.full},
	doi = {10.1117/12.2561087},
	eventtitle = {Space Telescopes and Instrumentation 2020: Optical, Infrared, and Millimeter Wave},
	pages = {314--323},
	booktitle = {Proceedings of the {SPIE}},
	publisher = {{SPIE}},
	author = {Zhou, Hanying and Krist, John and Seo, Byoung-Joon and Kern, Brian and Cady, Eric and Poberezhskiy, Ilya},
	urldate = {2024-12-30},
	date = {2020-12-13},
	year = {2020},
	month = {12},
}

@article{stetson_daophot_1987,
	title = {{DAOPHOT}: A Computer Program for Crowded-Field Stellar Photometry},
	volume = {99},
	issn = {0004-6280},
	url = {https://ui.adsabs.harvard.edu/abs/1987PASP...99..191S},
	doi = {10.1086/131977},
	shorttitle = {{DAOPHOT}},
	pages = {191},
	journal = {Publications of the Astronomical Society of the Pacific},
	publisher = {{IOP}},
	author = {Stetson, Peter B.},
	urldate = {2026-06-18},
	date = {1987-03-01},
	note = {{ADS} Bibcode: 1987PASP...99..191S},
  year = {1987},
  month = {mar},
}

@article{hom_romanRDI_2026,
	title = {{CoronaGraph} Instrument Reference Stars for Exoplanets ({CorGI}-{REx}). I. Preliminary Vetting and Implications for the Roman Coronagraph and Habitable Worlds Observatory},
	volume = {171},
	issn = {0004-6256},
	url = {https://ui.adsabs.harvard.edu/abs/2026AJ....171...36H/abstract},
	doi = {10.3847/1538-3881/ae1d68},
	pages = {36},
	number = {1},
	journal= {The Astronomical Journal},
	author = {Hom, Justin and Wolff, Schuyler G. and Clark, Catherine A. and Ciardi, David R. and Deveny, Sarah J. and Howell, Steve B. and Greenbaum, Alexandra Z. and Littlefield, Colin and Anche, Ramya M. and Bailey, Vanessa P. and Brandner, Wolfgang and Chauvin, Gaël and Girard, Julien H. and Kern, Brian and Mamajek, Eric and Mennesson, Bertrand and Savransky, Dmitry and Stapelfeldt, Karl R. and Biller, Beth A. and Brinjikji, Marah and Kuzuhara, Masayuki and Millar-Blanchaer, Maxwell A. and Mizuki, Toshiyuki and Schragal, Nicholas T. and Vega-Pallauta, Macarena C. and Wang, Jason J. and De Rosa, Robert J. and Douglas, Ewan S. and Macintosh, Bruce and Zhang, Jingwen and Program, The Roman Coronagraph Community Participation},
	urldate = {2026-07-05},
	date = {2026-01},
	year = {2026},
	month = {01},
	langid = {english},
}

@article{harris2020NumPy,
  title={Array programming with {NumPy}},
  author={Harris, Charles R. and Millman, K. Jarrod and van der Walt, Stéfan J. and Gommers, Ralf and Virtanen, Pauli and Cournapeau, David and Wieser, Eric and Taylor, Julian and Berg, Sebastian and Smith, Nathaniel J. and Kern, Robert and Picus, Matt and Hoyer, Stephan and van Kerkwijk, Marten H. and others},
  journal={Nature},
  volume={585},
  number={7825},
  pages={357--362},
  year={2020},
  publisher={Nature Publishing Group},
  doi={10.1038/s41586-020-2649-2}
}

@article{hunter2007matplotlib,
  title={Matplotlib: A 2D graphics environment},
  author={Hunter, John D.},
  journal={Computing in Science \& Engineering},
  volume={9},
  number={3},
  pages={90--95},
  year={2007},
  publisher={IEEE},
  doi={10.1109/MCSE.2007.55}
}

@article{virtanen2020scipy,
  title={SciPy 1.0: Fundamental Algorithms for Scientific Computing in Python},
  author={Virtanen, Pauli and Gommers, Ralf and Oliphant, Travis E. and Haberland, Matt and Reddy, Tyler and Cournapeau, David and Burovski, Evgeni and Peterson, Pearu and Weckesser, Warren and Bright, Jonathan and van der Walt, Stéfan J. and Brett, Matthew and Wilson, Joshua and Millman, K. Jarrod and others},
  journal={Nature Methods},
  volume={17},
  number={3},
  pages={261--272},
  year={2020},
  publisher={Nature Publishing Group},
  doi={10.1038/s41592-019-0686-2}
}

@inproceedings{lundberg2017shap,
	location = {Red Hook, {NY}, {USA}},
	title = {A unified approach to interpreting model predictions},
	isbn = {978-1-5108-6096-4},
	url = {https://dl.acm.org/doi/10.5555/3295222.3295230},
	series = {{NIPS}'17},
	pages = {4768--4777},
	booktitle = {Proceedings of the 31st International Conference on Neural Information Processing Systems},
	publisher = {Curran Associates Inc.},
	author = {Lundberg, Scott M. and Lee, Su-In},
	urldate = {2026-07-08},
	year = {2017},
}
\bibliographystyle{spiebib} 

\section{Appendix}

\subsection{Feature Selection}
\label{app:features}

Table~\ref{tab:selected_features} details the 12 features retained after the Iterative Feature Elimination (IFE) process. To formalize the mathematical definitions, let $\mu_t$ and $\sigma_t$ denote the temporal mean and standard deviation computed pixel-wise across the channel axis (comprising all unprobed and probed images). 

For the spatial dimension, the functions $\mu_s(\cdot)$ and $\sigma_s(\cdot)$ denote the local spatial mean and standard deviation, respectively. These are computed over the full images using the sliding spatial kernel of size $N_k \times N_k$, which is specifically scaled to encompass the expected Point Spread Function (PSF) core and incorporate the necessary local spatial context. Furthermore, let $I_0$ represent the original unprobed baseline scene, and $I_i^+$ denote the image modulated by the $i$-th positive probe.

We used SHAP (SHapley Additive exPlanations)\cite{lundberg2017shap} to rank the features. For each experiment (type of coronagraph, number of probes and amplitude of the PWP probes), it shows the dominant discriminators between true planet candidates and background speckles. Fig.~\ref{fig:shap_feature_imp} shows this ranking for  the perfect coronagraph using 3 probes with a probe amplitude of amplitude $\frac{\lambda}{16}$.

\begin{figure}[!ht]
    \centering
    \includegraphics[width=0.7\textwidth]{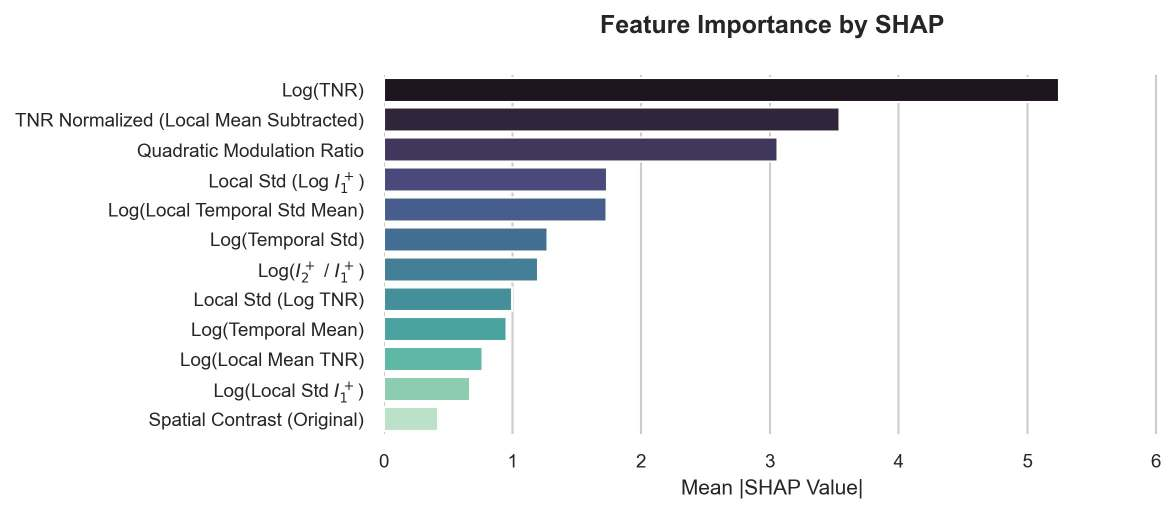}
    \caption{\textbf{Global feature importance for the \texttt{XGBoost} classifier evaluated via SHAP (SHapley Additive exPlanations).} The bar chart ranks the engineered features by their mean absolute SHAP values, illustrating their relative impact on the model's final binary predictions. This specific evaluation corresponds to the configuration for the perfect coronagraph using 3 probes with an amplitude = $\frac{\lambda}{16}$. The analysis reveals that the logarithmic Temporal Noise Ratio (Log(TNR)) and its locally normalized variant are the most dominant discriminators between true planet candidates and background speckles, followed by the Quadratic Modulation Ratio and local statistical metrics.}
    \label{fig:shap_feature_imp} 
\end{figure}

\begin{table}[htbp]
\centering
\small
\renewcommand{\arraystretch}{1.5}
\begin{tabular}{|p{4.5cm}|c|p{7cm}|}
\hline
\textbf{Feature} & \textbf{Math Definition} & \textbf{Description} \\ \hline
\multicolumn{3}{|c|}{\textbf{Temporal Statistics}} \\ \hline
Log(Temporal Mean) & $\log_{10}(\mu_t)$ & Captures the overall average intensity of the pixel across all probe states. \\ \hline
Log(Temporal Std) & $\log_{10}(\sigma_t)$ & Captures the raw temporal variability. \\ \hline
Log(TNR) & $\log_{10}(\sigma_t / \mu_t)$ & Normalizes variability by the mean to highlight coherent speckle modulation over static background. \\ \hline

\multicolumn{3}{|c|}{\textbf{Local Spatial-Temporal Variations}} \\ \hline
Log(Local Temporal Std Mean) & $\log_{10}(\mu_s(\sigma_t))$ & Local spatial mean of the temporal standard deviation. \\ \hline
TNR Normalized (Local Mean Subtracted) & $\log_{10}(\text{TNR}) - \mu_s(\log_{10}(\text{TNR}))$ & Log(TNR) mean-subtracted by its local spatial neighborhood, highlighting highly localized anomalous temporal behaviors. \\ \hline
Log(Local Mean TNR) & $\log_{10}(\mu_s(\text{TNR}))$ & Logarithm of the local spatial mean of the TNR. \\ \hline
Local Std (Log TNR) & $\sigma_s(\log_{10}(\text{TNR}))$ & Identifies spatial regions with highly fluctuating temporal noise. \\ \hline

\multicolumn{3}{|c|}{\textbf{Probe Modulation and Contrast}} \\ \hline
Quadratic Modulation Ratio & $\log_{10}\left( \frac{\sum (I_i^+ - I_0)^2}{I_0^2} \right)$ & Logarithmic ratio of the modulated power induced by the probes relative to the unprobed baseline. Quantifies the structural response to the focal plane probes. \\ \hline
Log($I_2^+ / I_1^+$) & $\log_{10}(I_2^+ / I_1^+)$ & Logarithmic intensity ratio between the second and first positive probe states. \\ \hline

\multicolumn{3}{|c|}{\textbf{Baseline Spatial Properties}} \\ \hline
Spatial Contrast ($I_0$) & $(I_0 - \mu_s(I_0)) / (\sigma_s(I_0) + \epsilon)$ & Spatial contrast of the  unprobed baseline image. \\ \hline
Log(Local Std $I_1^+$) & $\log_{10}(\sigma_s(I_1^+))$ & Logarithm of the local spatial standard deviation of the first probed state. \\ \hline
Local Std (Log $I_1^+$) & $\sigma_s(\log_{10}(I_1^+))$ & Local spatial standard deviation of the logarithm of $I_1^+$. \\ \hline
\end{tabular}
\caption{Mathematical definitions and physical descriptions of the 12 selected  features extracted from the full-frame $N \times N$ field of view.}
\label{tab:selected_features}
\end{table}


\subsection{Classification Evaluation Metrics}
\label{app:metrics}

To formalize the evaluation metrics used to assess the algorithm's performance, we map the binary classifications into a standard confusion matrix. In our specific architectural context, the classification outcomes are defined as follows:

\begin{itemize}
    \item \textbf{True Positives ($TP$):} Feature patches centered on an actual injected planet that are correctly classified as planets (label 1).
    \item \textbf{False Positives ($FP$):} Feature patches centered on unassociated speckles or background noise that are incorrectly classified as planets (label 1).
    \item \textbf{False Negatives ($FN$):} Feature patches centered on an actual injected planet that the algorithm incorrectly classifies as noise (label 0).
    \item \textbf{True Negatives ($TN$):} Feature patches centered on speckles or background noise that are correctly classified as noise (label 0).
\end{itemize}

Based on these foundational quantities, we define:

\begin{equation}
    \text{Precision} = \frac{TP}{TP + FP}
\end{equation}

\begin{equation}
    \text{Recall} = \frac{TP}{TP + FN}
\end{equation}

Consequently, the $F_\beta$ score, which balances both metrics under a customizable weighting factor $\beta$, is formulated as:

\begin{equation}
    F_\beta = (1 + \beta^2) \frac{\text{Precision} \cdot \text{Recall}}{(\beta^2 \cdot \text{Precision}) + \text{Recall}} = \frac{(1 + \beta^2) \cdot TP}{(1 + \beta^2) \cdot TP + \beta^2 \cdot FN + FP}
\end{equation}

\noindent Setting $\beta = 0.5$ yields the $F_{0.5}$ score, mathematically demonstrating how the metric penalizes False Positives ($FP$) twice as severely as False Negatives ($FN$), aligning the evaluation with our strict observational constraints.

\end{document}